\documentstyle[psfig]{aipproc}
\newcommand{\gpm}[3]{$#1^{+#2}_{-#3}$}

\begin{document}
\title{The decay of optical emission from the $\gamma$-ray burst GRB 970228}

\author{Titus J. Galama$^*$, Paul J. Groot$^*$, Jan van
Paradijs$^{*\dagger}$, Chryssa Kouveliotou$^{\ddagger \|}$,
Kailash C. Sahu$^{\oplus}$, Mario Livio$^{\oplus}$, Larry
Petro$^{\oplus}$, F. Duccio 
Macchetto$^{\oplus}$ and Andrew Fruchter$^{\oplus}$}
\address{$^*$ Astronomical Institute ``Anton Pannekoek'', University
of Amsterdam, \& Center for High Energy Astrophysics, Kruislaan 403,
1098 SJ Amsterdam, The Netherlands.\\
$^{\dagger}$ Physics Department, University of Alabama in Huntsville,
Huntsville, AL  
35899, USA \\
$\ddagger$ Universities Space Research Association \\
$\|$ NASA Marshall Space Flight Center, ES-84, Huntsville, AL 35812, 
USA\\
$\oplus$ Space Telescope Science Institute, 3700 San Martin Drive,
Baltimore, MD 21218, USA}

\maketitle

\begin{abstract}
We present the R$_{\rm c}$ band light curve 
of the optical transient (OT)
associated with GRB970228, based on re-evaluation of existing
photometry. Data obtained until April 1997 suggested a
slowing down of the decay of the optical brightness. However, the HST
observations in September 1997 show that the light curve of the point
source is well represented by a single power law, with a ``dip'',
about a week after the burst occured. The exponent of the power law
decay is $\alpha$ =
--1.10 $\pm$ 0.04. As the
point source weakened it also became redder. 
\end{abstract}

\section*{Introduction}

The $\gamma$-ray burst of February 28, 1997, detected\cite{Costa_a} with the 
Gamma-Ray Burst Monitor on the BeppoSAX observatory, and located with an $\sim 
3^{\prime}$ radius position with the Wide Field Camera on the same satellite,
was the first for which a fading X-ray\cite{Costa_a} and optical
counterpart\cite{Groot_a,jvp_a}  were found. 


The optical counterpart was discovered from a
comparison of V and I$_{\rm c}$ 
band images taken with the William Herschel Telescope (WHT) on February 28.99
UT, and the Isaac Newton Telescope (INT; V  band) and the WHT (I$_{\rm
c}$  band) on March 
8.86 UT. After the counterpart had weakened by 
several magnitudes, it was found
to coincide with an extended object\cite{jvp_a,Groot_b,Metzger}. 
In subsequent observations with
the Hubble Space Telescope (HST) on March 26 and April 7, 1997, it 
was found that the optical counterpart consists of a point source and
an extended ($\sim 1^{\prime \prime}$) object, offset from the point source by
$\sim 0.5^{\prime \prime}$\cite{Sahu_a}. 

We here reassess the photometric information presented by Galama et al. 
\cite{Galama}
in the light of the recent HST findings\cite{Fruchter,Fruchter2},
and present the R$_{\rm c}$ band optical light
curve of the GRB counterpart.

\section{Observations}
In Table \ref{Data1} we have collected the optical photometry reported 
on GRB970228, obtained in the V,
R$_{\rm c}$, and
I$_{\rm c}$ passbands (effective wavelengths $\sim 5500  , \sim 6500$,
and $\sim 
8000$ \AA, 
respectively, corresponding closely to the Cousins VRI system). In
the interpolations to the R$_{\rm c}$ band (see Table
\ref{Data2}) we have assumed
that the spectra of both the point 
source and the extended emission are smooth (i.e., not dominated by
emission lines). We have used the
relation between the color indices V-R$_{\rm c}$ and V-I$_{\rm c}$
given by Th\'e et 
al.\cite{The} for late-type stars; for bluer stars we have inferred
this relation 
from the tables given by Johnson\cite{Johnson} for main-sequence stars
and the color 
transformations to the Cousins VRI system given by Bessel\cite{Bessel}.
We have tested the validity of these color-color relations from
numerical integrations of power law flux distributions and
of Planck functions,
and conclude that if the flux distribution of the optical counterpart
is smooth, the uncertainty in the interpolated R$_{\rm c}$ magnitude
is unlikely to exceed 0.1 magnitude\cite{Galama}.  
Here we discuss the differences with respect to Galama et al.\cite{Galama}. \\

\begin{figure}[b!] 
\centerline{\psfig{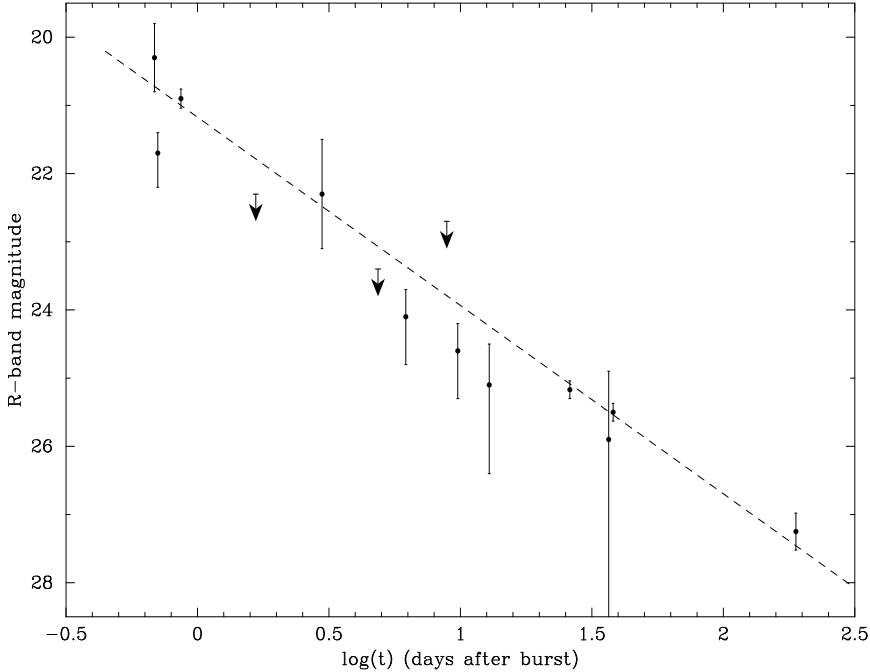}}
\vspace{10pt}
\caption{The R$_{\rm c}$
band lightcurve of GRB 970228. Indicated is a power-law
fit, $F_{\nu} \propto \nu^{\alpha}$, to the data with slope $\alpha$ =
--1.10 $\pm$ 0.04 ($\chi^{2}_{\rm r}$ = 2.3 with 9 degrees of freedom).}
\label{Lightcurve}
\end{figure}

The earliest image of the OT was obtained by
Pedichini et al.\cite{Pedichini}. This observation was obtained with a
wide filter; we have transformed this wide filter magnitude, using the
reported filter characteristics, to the
R$_{\rm c}$ band. 
In the images of Guarnieri et
al. \cite{Guarnieri}, the OT is blended with the nearby late type
star, due 
to bad seeing. We corrected for the contribution of the 
late type star (for which  
$R_{\rm c}$ = 22.1 $\pm$ 0.1\cite{Galama}) and find for the OT $R_{\rm c}$ = 
\gpm{21.7}{0.5}{0.3} (Note that in \cite{Galama} the time of the
observation of Guarnieri et al.\cite{Guarnieri} is given slightly
incorrect). We 
have included a lower limit to the R$_{\rm c}$  
magnitude on Mar. 1.791\cite{Guarnieri}, which after correction 
for the contribution of the late type star, gives $R_{\rm c}$ $>$ 22.2.
We have not included the subsequent measurements and upper-limits as given by
Guarnieri et al. \cite{Guarnieri} as they are consistent with detections of
the late-type star only. We noticed that the Keck
calibration\cite{Metzger,Metzger_b} differs
by +0.3 magnitudes; we
have corrected for this in Table \ref{Data2}.   
HST STIS observations between             
Sept. 4.65 and 4.76 UT \cite{Fruchter,Fruchter2} show that both 
the nebula and the point-source optical   
transient found in earlier HST             
WFC2 observations\cite{Sahu_a},  are also detected in the STIS
images at a level of  $V$ = 28.0 $\pm$ 0.25 for the point source and
$V$ = 25.7 $\pm$ 0.15  
for the nebula. Reanalysis of the earlier HST WFPC2 
F606W observations gives $V$ = 25.6 $\pm$ 0.25 for the nebula\cite{Fruchter}.
From the recent HST observations and the reanalysis we infer that the
R$_{\rm c}$ band magnitude of the nebula is $R_{\rm c}$ = 25.0 $\pm$
0.3, fainter than but 
consistent with $R_{\rm c}$=24.7 $\pm$ 0.3\cite{Galama}.
Assuming that the colors of the OT remained constant during
the decay (i.e., taking the observed $V-I$ = 1.85 from the HST March 26 and
April 7 observations) we infer from the Sept. 4.71 HST observations 
that $R_{\rm c}$ = 27.25 $\pm$ 0.27 for the OT.
In all ground-based photometry we corrected for the contribution of
the extended object ($R_{\rm c}$ = 25.0  $\pm$ 0.3) and show the
results in Table \ref{Data2} and in Fig. \ref{Lightcurve}. 
We have fitted a power-law, $F_{R} = F_{0}t^{\alpha}$,
 to the detections and find a magnitude $m_{0}$ = 21.14 $\pm$
0.13 (corresponding to  $F_{0}$) and a slope $\alpha$ =
--1.10 $\pm$ 0.04 ($\chi^{2}_{\rm r}$ = 2.3; the three upper-limits
are not included in this fit).
\section{Discussion}

The lightcurve can be well represented by a power-law. In the interval
between log$(t)$ = 0.7 and 1.2 we have three detections and one
upper-limit located below the power-law fit. The three detections
deviate from  the power-law by:  1.9 $\sigma$ for the Keck
observation (Mar 6.32 UT), 1.7 $\sigma$ for the INT
observation (Mar 9.90 UT) and 1.4 $\sigma$ for the NTT observation
(Mar 13.00 UT).
This might indicate that the lightcurve is not smooth, but superposed on the
power law behaviour we have deviations of small amplitude. A similar
result has been found for the OT of GRB 970508\cite{Galama_b,Groot}.
As the
point source weakened it also became redder ($V-I$ = 0.7
$\pm$ 0.14 on Feb. 28.99 to $V-I$ = 1.90 $\pm$ 0.14 on March 26 and 
$V-I$ = 1.80 $\pm$ 0.14 on April 7; see Tab. \ref{Data1}).

\begin{table}
\caption[ ]{Summary of optical observations. \label{Data1}}
\begin{tabular}{llll}
Date (UT) & \multicolumn{1}{c}{Telescope\tablenote{Abbreviations: RAO,
Rome Astrophysical Observatory; BUT, Bologna University  Telescope;
WHT, William Hershell Telescope; APO, Apache Point Observatory; NOT,
Nordic Optical Telescope, INT, Isaac Newton Telescope; HST, Hubble
Space Telescope; P5m, Palomar 5-m Hale telescope.}} & 
Magnitude& \multicolumn{1}{c}{Remarks\tablenote{Abbreviations: OT,
optical transient; EXT, extended source; LTS, late-type star.}} 
\\
\tableline
Feb. 28.81 & RAO & wide $R=20.5 \pm 0.5$ & OT \\
Feb. 28.83 & BUT & $R=21.1 \pm 0.2$ & OT+LTS\\
Feb. 28.99 & WHT & $V=21.3 \pm 0.1$ & OT \\
Feb. 28.99 & WHT & $I=20.6 \pm 0.1$& OT\\
Mar. 01.79 & BUT & $R>21.4$ & OT+LTS\\
Mar. 03.10 & APO & $B=23.3 \pm 0.5$& OT\\
Mar. 04.86 & NOT & $V>24.2$ & OT+EXT\\
Mar. 06.32 & Keck & $R=24.0$ & OT+EXT \\
Mar. 08.86 & INT & $V>23.6$ & OT+EXT\\
Mar. 08.88 & WHT & $I>22.2$ & OT+EXT\\
Mar. 09.85 & INT & $B=25.4 \pm 0.4$ & OT+EXT\\
Mar. 09.90 & INT & $R=24.0 \pm 0.2$ & OT+EXT\\
Mar. 13.00 & NTT & $R=24.3 \pm 0.2$ & OT+EXT\\
Mar. 26.38 & HST & $V=26.1 \pm 0.1$ & OT\\
Mar. 26.47 & HST & $I=24.2 \pm 0.1$ & OT\\
Mar. 26.38 \& Apr. 07.22 & HST & $V=25.6 \pm 0.25$ & EXT\\
Mar. 26.47 & HST & $I=24.5 \pm 0.3$ & EXT\\
Apr. 05.76 & Keck& $R=24.9 \pm$ 0.3 & OT+EXT\\ 
Apr. 07.22 & HST & $V=26.4 \pm 0.1$ & OT\\
Apr. 07.30 & HST & $I=24.6 \pm 0.1$ & OT\\
Apr. 07.30 & HST & $I=24.3 \pm 0.35$ & EXT\\
Sept 04    & P5m & $R=25.5 \pm 0.5$ & OT+EXT\\
Sept 04.71 & HST & $V=28.0 \pm 0.25$ & OT\\
Sept 04.71 & HST & $V=25.7 \pm 0.15$ & EXT\\
\tableline
\end{tabular}
\end{table}

\begin{table}
\caption[ ]{The R-band lightcurve of GRB 970228.
\label{Data2}}
\begin{tabular}{llllll}
Date(UT) & \multicolumn{1}{c}{Telescope\tablenote{Abbreviations as in
Table \ref{Data1}.}} & R(OT+LTS+EXT) & R(OT+EXT) & R(OT) & Reference\\
\tableline
Feb. 28.81 & RAO & & 20.3 $\pm$ 0.5 & 20.5 $\pm$ 0.5 & \cite{Pedichini} \\
Feb. 28.83 & BUT & 21.1 $\pm$ 0.2 & \gpm{21.7}{0.5}{0.3} &
\gpm{21.7}{0.5}{0.3} & \cite{Guarnieri} \\
Feb. 28.99 & WHT & &20.9 $\pm$ 0.14&  20.9 $\pm$ 0.14 & \cite{Galama}\\
Mar. 01.79 & BUT & $>21.4$ & $>22.2$ & $>22.3$ & \cite{Guarnieri}\\
Mar. 03.10 & APO & &22.2 $\pm$ 0.7 &  \gpm{22.3}{0.8}{0.8} & \cite{Galama} \\
Mar. 04.86 & NOT & &$>$23.3  & $>$23.4 & \cite{Galama}\\
Mar. 06.32 & Keck& &23.7 $\pm 0.2$ &  \gpm{24.1}{0.5}{0.4}
&\cite{Metzger} \\
Mar. 08.88 & INT+WHT & &$>$22.6  & $>$22.7 &\cite{Galama}\\
Mar. 09.90 & INT & &24.0 $\pm$ 0.2 &  \gpm{24.6}{0.7}{0.4} &\cite{Galama}\\
Mar. 13.00 & NTT & &24.3 $\pm$ 0.2 &  \gpm{25.1}{1.3}{0.6} &\cite{Galama}\\
Mar. 26.20 & HST & &24.3 $\pm$ 0.2  &  25.17 $\pm$ 0.13 &\cite{Galama,Sahu_a}\\
Apr. 05.76 & Keck& &24.6 $\pm$ 0.3 & \gpm{25.9}{\infty}{1.0} & \cite{Metzger_b} \\ 
Apr. 07.23 & HST & &24.5 $\pm$ 0.15 &  25.50 $\pm$ 0.13 &\cite{Galama,Sahu_a}\\
Sep. 04    & P5m & &25.5 $\pm$ 0.5 & & \cite{Djorgovski} \\
Sep  04.71 & HST & &24.9 $\pm$ 0.3  &27.25 $\pm$ 0.27 & \cite{Fruchter,Fruchter2}\\
\tableline
\end{tabular}
\end{table}


\begin{references}
\bibitem{Costa_a} Costa, E. et al., {\it Nat}\ {\bf 387}, 783
(1997)
\bibitem{Groot_a} Groot, P.J. et al., {\it IAU Circular} No. 6584 (1997). 
\bibitem{jvp_a} Van Paradijs, J. et al., {\it Nat}\ {\bf 386}, 686 (1997) 
\bibitem{Groot_b} Groot, P.J. et al., {\it IAU Circular} No. 6588 (1997). 
\bibitem{Metzger} Metzger, M.R. et al., {\it IAU Circular} No. 6588
(1997). 
\bibitem{Sahu_a} Sahu, K. et al., {\it Nat}\ {\bf 387}, 476 (1997)
\bibitem{Galama} Galama, T.J. et al., {\it Nat}\ {\bf 387}, 479 (1997)
\bibitem{Fruchter} Fruchter et al., {\it IAU Circular} No. 6747 (1997). 
\bibitem{Fruchter2} Fruchter et al., {\it These proceedings} (1997). 
\bibitem{The} Th\'e, P.S., Steenman, H., Alcaino, G., {\it A\&A}\ {\bf
132}, 385 (1984)
\bibitem{Johnson} Johnson, H.L., {\it ARA\&A}\ {\bf 4}, 191 (1966)
\bibitem{Bessel} Bessel, M.S., {\it UBVRI Photometry with a Ga-As
Photomultiplier, PASP} {\bf 88}, 557 (1976) 
\bibitem{Pedichini} Pedichini et al., {\it A\&A}\ {\bf 327}, L32 (1997)
\bibitem{Guarnieri} Guarnieri, A. et al., {\it submitted to A\&A}\
Astro-ph 9707164 (1997) 
\bibitem{Metzger_b} Metzger, M.R. et al., {\it IAU Circular} No. 6631 (1997). 
\bibitem{Galama_b} Galama, T.J. et al., {\it submitted to ApJL}\
(1997)
\bibitem{Groot} Groot, P.J. et al., {\it These proceedings}\ (1997)
\bibitem{Djorgovski} Djorgovski, S.G., et al.,  {\it IAU Circular} No. 6732
(1997).   

\end{references}
\end{document}